\documentclass[twocolumn,superscriptaddress,groupedaddress,amsmath,amssymb,aps,prl,floatfix]{revtex4-1}

\usepackage{comment}
\usepackage{graphicx}
\usepackage{dcolumn}
\usepackage{bm}

\usepackage[utf8]{inputenc}
\usepackage[T1]{fontenc}

\usepackage{braket}
\usepackage{siunitx}
\usepackage{xcolor}

\usepackage[normalem]{ulem}

\def\Rb87{^{87}\mathrm{Rb}}                             

\begin{document}

\title{Turbulence in the Geomagnetic Field at Earth Surface}

\author{Mingshu Zhao}
\affiliation{Hebei Key Laboratory of Physics and Energy Technology, Department of Mathematics and Physics, North China Electric Power University, Baoding, Hebei 071003, China}
\email{zmshum@ncepu.edu.cn}
\author{Xiaoping Zeng}
\affiliation{Institute of Geophysics, China Earthquake Administration, Beijing 100081, China}
\author{Yunfang Lin}
\affiliation{Institute of Geophysics, China Earthquake Administration, Beijing 100081, China}

\date{\today} 

\begin{abstract}
We investigated turbulence-like behavior in the geomagnetic field using ground-based magnetic observatory data across China. Through analysis of spatial and temporal structure functions, we find power-law scaling consistent with Kolmogorov-like turbulence under extended self-similarity. We also identify significant correlations between vertical geomagnetic field variations and large earthquakes. The combination of turbulent characteristics with these correlations suggests a physical mechanism where solar activity provides energy that is transferred through turbulent processes to smaller scales, potentially contributing to earthquake triggering.
\end{abstract}

\maketitle

Earth’s geomagnetic field serves as a critical shield, protecting the planet from charged particles in the solar wind while preserving the atmosphere necessary for life~\cite{kivelson1995introduction}. 
This protective role highlights a clear connection between the geomagnetic field and solar activity, as short-term variations in the field are predominantly driven by solar phenomena such as solar flares and coronal mass ejections~\cite{basavaiah2012geomagnetism, mandea2019geomagnetism}. 
Additionally, geophysical events like earthquakes and volcanic activity, though rare, can influence short-term geomagnetic variations, particularly in the vertical component, through mechanisms such as the piezomagnetic effect~\cite{stacey1964seismomagnetic, stacey1972theory, johnston2002electromagnetic} and electrokinetic effect~\cite{fitterman1979theory, johnston2002electromagnetic}.  Consequently, anomalous geomagnetic signals are increasingly studied as potential precursors to seismic events~\cite{zeng1998manual, ding2004geomagnetic,rabeh2010strong}.

While the geomagnetic field connects to both solar activity and earthquakes, establishing a direct correlation between solar activity and seismic events remains controversial~\cite{love2013insignificant,odintsov2006long,huzaimy2011possible,han2004possible,gribbin1971relation}. However, recent satellite observations provide compelling evidence of a correlation between solar activity and large earthquakes worldwide~\cite{marchitelli2020correlation}, suggesting that solar energy may play a role in triggering seismic events~\cite{simpson1967solar}.
The underlying physical processes, however, remain unresolved. 
In this letter, we propose a turbulence-mediated mechanism to explain how solar-driven geomagnetic disturbances could influence seismic activity. 
By modeling the geomagnetic field as a turbulent medium that transmits energy from solar disturbances to the Earth, we aim to bridge the gap between these seemingly independent phenomena.

Turbulence, originally developed to describe chaotic fluid motions, has provided explanations for diverse phenomena including cosmological inflation relying on gravitational wave turbulence~\cite{galtier2020plausible,GW2021turbulence} and primordial black hole formation during magneto-hydrodynamic (MHD) turbulence~\cite{liang2025primordial}. 
These successful applications across vastly different systems demonstrate turbulence's remarkable versatility in modeling energy transfer processes. 
Building on this foundation, we apply the concept of direct energy cascade~\cite{kolmogorov1995turbulence} in turbulence to explain the solar-earthquake correlation. 

The turbulent energy cascade involves large-scale injection, inertial-range transfer (exhibiting self-similar scaling), and small-scale dissipation. 
In the Sun-Earth system, solar activities inject energy at large scales, which cascades through MHD turbulence to smaller scales. 
Ultimately, this energy is dissipated at Earth’s small scales through complex processes that may contribute to earthquake triggering.

We validate our model by testing for turbulence signatures in geomagnetic field structures.
While previous studies have characterized temporal properties of geomagnetic turbulence~\cite{de2004time}, its spatial structure—particularly at ground level—remains largely unexplored. Turbulent features have been observed in the magnetosphere extending several tens of thousands of kilometers from Earth’s surface~\cite{borovsky2003role,zimbardo2006magnetic}. However, these signatures attenuate substantially through the ionosphere and atmosphere before reaching the surface~\cite{mandea2019geomagnetism}, leaving open whether turbulence-like spatial structures persist in ground-based measurements.

In this work, we analyze ground-based magnetic field data from observatories across China to investigate the statistical properties of short-term geomagnetic variations. 
Using structure function analysis~\cite{kolmogorov1995turbulence}, we identify power-law scaling behaviors under the extended self-similarity (ESS)~\cite{benzi1993extended}, revealing turbulence-like spatial and temporal structures similar to those observed in fluid turbulence. 
Our analysis also reveals an intriguing correlation between seismic activity and solar influences. 
We find large earthquakes correlate with enhanced daily fluctuations in the vertical geomagnetic component, suggesting strong solar-terrestrial coupling. Together with observed turbulence properties, this indicates solar energy may cascade through the geomagnetic field to ultimately trigger seismic events.

\begin{figure*}[t]
\includegraphics{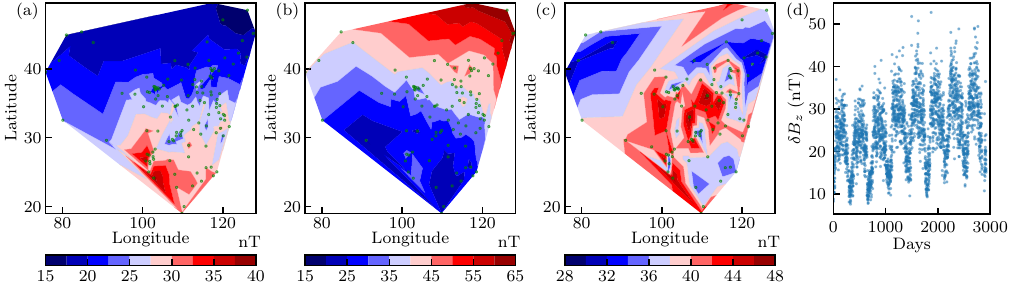}
\caption{Daily variation of vertical geomagnetic field $\delta B_z$. 
(a) 8-year mean $\delta B_z$ contour over China. 
(b) $\delta B_z$ contour on 16 July 2012. 
(c) $\delta B_z$ contour on 5 April 2010. 
Green points indicate observatory locations. 
(d) Time series of daily $\delta B_z$ averaged across all stations (8-year period).
}
\label{fig:contour}
\end{figure*}

{\it Dataset}
The dataset comprises daily changes in the vertical ($Z$) component of the geomagnetic field, denoted $\delta B_z$, collected from approximately 100 ground-based observatories across China from 1 February, 2008, to 1 February, 2016, with measurements recorded at one-minute intervals.
The $Z$ component better reflects seismic activity while being less disturbed by geomagnetic storms~\cite{zeng1998manual,ding2004geomagnetic}.

We analyze $\delta B_z$ instead of absolute values $B_z$ to highlight short-term fluctuations while avoiding inconsistencies in reference values across observatories. This choice aligns with Reynolds decomposition~\cite{kolmogorov1995turbulence}, treating $\delta B_z$ as the fluctuation component relevant for turbulence analysis. 
The observed $\delta B_z$ typically range from $15$ to $40\ \rm{nT}$, as illustrated in Fig.~\ref{fig:contour}(a), which shows the contour of the mean $\delta B_z$ over the 8-year period. While the mean $\delta B_z$ predominantly exhibits a latitudinal dependence—with higher latitudes generally corresponding to smaller $\delta B_z$ values—this simple spatial pattern is not consistently observed on individual days. For instance, Fig.~\ref{fig:contour}(b) displays a realization with an opposite latitudinal trend, whereas Fig.~\ref{fig:contour}(c) reveals a more complex spatial structure.

Figure.~\ref{fig:contour}(d) presents the time series of $\delta B_z$ averaging all stations each day in $8$ years. Despite the presence of long-term drift, we observe a clear oscillation with an approximate one-year period, suggesting that $\delta B_z$ is largely influenced by Earth's revolution around the Sun, which modulates solar wind-magnetosphere coupling efficiency.


{\it Spatial Structures}
To investigate the spatial structure of geomagnetic field fluctuations across China, we define the spatial increments of $\delta B_z$ between two observatories $i$ and $j$, separated by a great-circle distance $\Delta x$, as 
$\Delta B_z(\Delta x) = \bigl|\delta B_{z,i} - \delta B_{z,j}\bigr|$, 
where the subscripts $i$ and $j$ denote the two observatories. 
$\Delta x$ is computed from the two sites’ latitude and longitude using the Haversine 
formula~\cite{van2017heavenly}, with Earth’s radius $R = 6371\,\mathrm{km}$.

The spatial structure function $S_p^L(\Delta x)$ is defined in terms of the spatial increments 
$\Delta B_z(\Delta x)$ as 
$S_p^L(\Delta x) = \langle |\Delta B_z(\Delta x)|^p \rangle$, 
where $p$ is the order of the structure function, and $\langle \cdots \rangle$ represents 
the average over all data points with separation $\Delta x$. 
To facilitate statistical analysis, we discretize $\Delta x$ into $18$ logarithmically spaced bins 
ranging from $110\,\mathrm{km}$ to $2430\,\mathrm{km}$, grouping together data 
with similar separation distances.

For systems whose energy can be decomposed across different scales, self-similarity often emerges, 
leading to power-law scaling of the structure function~\cite{aluie2011compressible,aluie2013scale}, i.e., 
$S_p(\Delta x) \propto (\Delta x)^{\zeta_p}$. 
Under the assumptions of isotropy, homogeneity, and high Reynolds number, 
the K41 model predicts $\zeta_p = p/3$ for incompressible fluid turbulence~\cite{kolmogorov1991local, kolmogorov1941degeneration, kolmogorov1991dissipation}. 
However, at moderate Reynolds numbers, the available scaling range is limited. 
ESS addresses this limitation by directly comparing structure functions of different orders, which helps recover the scaling~\cite{benzi1993extended,benzi1993extended2}. 
When compared to $S_3$, ESS reproduces the K41 prediction and has been observed in many geophysical systems~\cite{carbone1996evidences,nikora2001extended}. 
Here we denote the ESS scalings by ~$\zeta^{\mathrm{ESS}}_p$ satisfying $S_p(\Delta x)\propto\left[S_3(\Delta x)\right]^{\zeta^{\mathrm{ESS}}_p}$.
Deviations from K41 scalings are frequently captured by the KO62 model~\cite{kolmogorov1962refinement,obukhov1962some}, which states that $\zeta^{\mathrm{ESS}}_p - p/3 = -\chi \, p \, (p - 3)$, 
where $\chi$ is the intermittency exponent.

In the near-surface geomagnetic system, it remains uncertain whether scale decomposition can be rigorously applied. 
Previous studies have reported non-Kolmogorov scalings in temporal structures~\cite{de2004time}. 
Additionally, the variations in latitude appear more pronounced than those in longitude, indicating that the fluctuations are inhomogeneous and anisotropic, as can be seen from Fig.~\ref{fig:contour}(a). As a result, we do not expect Kolmogorov’s homogeneous and isotropic turbulence theory to strictly apply. 
However, as a starting point, we still employ traditional turbulence analysis methods.
Furthermore, extended self-similarity may still be present, potentially arising from an inherent hierarchical structure~\cite{ching2002extended}.

\begin{figure}[tb]
\includegraphics{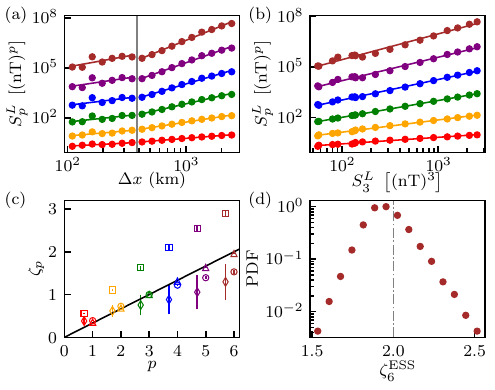}
\caption{Spatial structures.
(a) Spatial structure function $S_p^L$ for $p = 1\ldots6$ versus separation $\Delta x$. A vertical grey line at $L_0$ marks two scaling regimes. Lines show power-law fits. The color coding follows the x-axis scale of panel (c).
    (b) Extended self-similarity of $S_p^L$, with lines showing power-law fits.
    (c) Scaling exponents: diamonds and squares [from panel (a) for $L < L_0$ and $L > L_0$] and circles [from panel (b)], and triangles from the average of ESS scaling of each day. The black line is the K41 scaling $p/3$. The diamonds and squares are shifted left a bit to prevent overlap. 
    (d) PDF of $\zeta_6^{\rm ESS}$ over an 8-year period, normalized so its peak is 1. The grey dashed line marks the K41 scaling.
}
\label{fig:structure_functions}
\end{figure}
From our dataset, we identified two distinct scaling regimes with different power laws. Figure~\ref{fig:structure_functions}(a) presents $S_p^L(\Delta x)$, revealing two power-law behaviors separated at $L_0 \approx 390\,\mathrm{km}$ (grey line). In region~1 ($\Delta x < L_0$), the scaling exponent $\zeta^1_p$ is approximately $2/5$ of $\zeta^2_p$, the scaling in region~2 ($\Delta x > L_0$), as shown in Fig.~\ref{fig:structure_functions}(c). 
This suggests that at larger spatial separations, extreme values become more frequent due to underlying inhomogeneities—contrasting sharply with homogeneous classical turbulence, where smaller scales typically exhibit stronger intermittency. Consequently, to accurately extract the scaling behavior across the entire system, it is necessary to employ ESS, similar to previous studies on inhomogeneous turbulence in channel flows~\cite{toschi1999intermittency, arad1999disentangling}.

Figure~\ref{fig:structure_functions}(b) illustrates ESS by plotting $S_p^L$ against $S_3^L$, yielding the scaling exponents $\zeta_p^{\mathrm{ESS}}$ [circle markers in Fig.~\ref{fig:structure_functions}(c)]. These exponents closely follow the predictions of K41 but exhibit clear intermittency at $p>3$. A KO62 fit estimates an intermittency exponent of $\chi = 0.026(1)$.

To investigate the stability of ESS, we compute daily $\zeta^{\mathrm{ESS}}_p$ values and plot the probability density function (PDF) of $\zeta_6^{\mathrm{ESS}}$ in Fig.~\ref{fig:structure_functions}(d). The distribution deviates from a Gaussian form, with its mean [indicated by triangle markers in Fig.~\ref{fig:structure_functions}(c)] positioned closer to the K41 prediction than the circle markers, corresponding to an intermittency exponent of $\chi = 0.003(1)$. 
The daily ESS exhibits lower intermittency compared to the 8-year ESS, suggesting non-stationary behavior in the time series. This observation is consistent with Fig.~\ref{fig:contour}, which highlights significant day-to-day variations in the spatial structure.

Overall, our results indicate that geomagnetic field variations exhibit self-similarity and share structural characteristics with fluid turbulence, despite their inherent anisotropy and inhomogeneity.

\begin{figure}[t]
\includegraphics{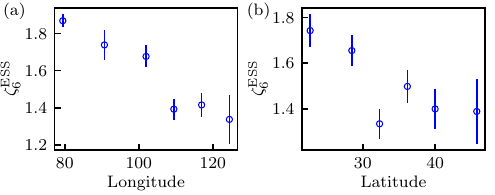}
\caption{
Inhomogeneity. ESS scaling exponent $\zeta_6^{\mathrm{ESS}}$ as a function of longitude (a) and latitude (b).
}

\label{fig:Inhomo}
\end{figure}

{\it Inhomogeneity}
We investigate inhomogeneity using a method similar to that in Ref.~\cite{toschi1999intermittency}. The longitude $\lambda$ is divided into six bins, and we compute $S_p^L(\lambda_i, \Delta x)$, which includes only spatial increments with at least one station within the $i$th longitude bin. The ESS scaling exponents $\zeta_6^{\mathrm{ESS}}$ are presented in Fig.~\ref{fig:Inhomo}(a), revealing clear inhomogeneity due to boundary effects. Specifically, higher $\lambda$ values—closer to the interface between the mainland and ocean—exhibit stronger intermittency.  
This boundary effect closely resembles the behavior observed in channel fluid flow~\cite{arad1999disentangling}.

We apply the same method to latitude $\phi$ by dividing it into bins and computing $S_p^L(\phi_i, \Delta x)$, including only spatial increments with at least one station within the $i$th latitude bin. The results, shown in Fig.~\ref{fig:Inhomo}(b), display a similar trend: higher latitudes exhibit stronger intermittency. 
This effect partially stems from the latitude-dependent $\delta B_z$ inhomogeneity shown in Fig.~\ref{fig:contour}(a).

We also apply the same method to study anisotropy. Despite a clear difference in amplitude—where the structure function in the latitude direction exhibits greater magnitude—the ESS scaling does not show a significant dependence on angle.

{\it Temporal Structures}
Similar to the spatial increments, we define the temporal increments of the 
time series $\delta B_z(t)$ as 
$\Delta B_z(\Delta t) = \delta B_z(t + \Delta t) - \delta B_z(t)$. 
The corresponding temporal structure function of order $p$ is then 
$S_p^T(\Delta t) = \langle |\Delta B_z(\Delta t)|^p \rangle$, 
where $\langle \cdots \rangle$ denotes an average over all data points separated by 
$\Delta t$. We refer to the exponent governing the temporal scaling of $S_p^T$ 
as $\tau_p$.

\begin{figure}[t]
\includegraphics{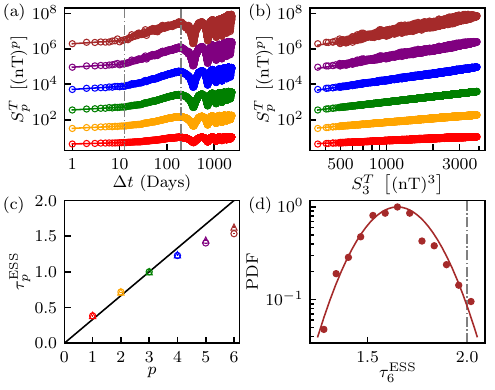}
\caption{
Temporal structure.
(a) Temporal structure functions $S_p^T(\Delta t)$ for $p = 1,\ldots,6$, plotted 
against the time separation $\Delta t$. The two vertical gray lines mark three distinct regimes; power-law fits are shown for the first two, while the 
third regime does not exhibit a clear power law.
(b) Extended self-similarity of $S_p^T$, with lines indicating power-law fits.
(c) Circles show the scalings from panel (b), while triangles present the station-averaged 
ESS scaling. The black line corresponds to the K41 scaling $p/3$.
(d) PDF of $\tau_6^{\rm ESS}$ for all stations, normalized so that 
its maximum is 1. The solid curve represents a Gaussian fit, and the gray dashed line 
indicates the K41 value.
}
\label{fig:time_series}
\end{figure}

To investigate the overall temporal structure across all stations, we compute the temporal structure function of the daily-averaged time series shown in Fig.~\ref{fig:contour}(d). 
The results, presented in Fig.~\ref{fig:time_series}(a), reveal three distinct regions separated by the gray dashed lines at $T_0 = 13$~days and $T_1 = 200$~days. 
In the third regime ($\Delta t > T_1$), an oscillatory pattern aligns with the $1$-year periodicity observed in the data. 
Although $S_p^T$ appears self-similar for different orders $p$, we find that the scaling in the first two regions saturates at $p \geq 4$, in contrast to the hierarchical structure seen for $p \leq 3$. 
While the third region does not exhibit a clear power law, it shows a more pronounced hierarchical structure. 
This behavior is also evident in Fig.~\ref{fig:time_series}(b), where $S_p^T$ is plotted against $S_3^T$ to verify ESS. 
The lines represent power-law fits, and the resulting exponents (circle markers) appear 
in Fig.~\ref{fig:time_series}(c). 
The intermittency observed here closely matches that of the spatial structure, with an intermittency exponent of $\chi = 0.0259(1)$.

To explore the stability of temporal ESS across stations, we compute the temporal structure function individually for each station. 
The PDF of the sixth-order ESS scaling exponent $\tau_6^{\rm ESS}$ is shown in Fig.~\ref{fig:time_series}(d). 
It follows a Gaussian distribution with $\mathrm{mean} = 1.63$ and $\sigma = 0.16$. 
The station-averaged result, indicated by triangle markers in Fig.~\ref{fig:time_series}(c), yields an intermittency exponent $\chi = 0.0212(5)$. 
Notably, these results closely resemble those in Fig.~\ref{fig:time_series}(b), which we attribute to the similarity in long-term structure across stations (all exhibiting a near one-year periodicity), despite differences in amplitude ranges.

In summary, our analysis reveals that the temporal structure of geomagnetic field fluctuations exhibits ESS and turbulence-like behavior, reinforcing the statistical properties observed in the spatial structure function analysis.

\begin{figure}[t]
\includegraphics{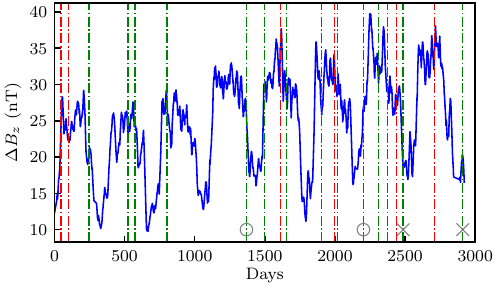}
\caption{
Earthquake. The smoothed $\delta B_z$ time series is shown in blue. 
Red vertical dashed lines mark earthquakes with magnitudes greater than $6.5$, 
while green vertical dashed lines correspond to magnitudes between $6.0$ and $6.5$. 
Gray crosses indicate events that do not coincide with a peak in the time series, 
and gray hollow circles mark earthquakes occurring during transitions between 
peaks and valleys.
}
\label{fig:earthquake}
\end{figure}

{\it Earthquakes}
Several studies have suggested that $\delta B_z$ are correlated with seismic activity and may serve as a precursor to earthquakes, an effect attributed to the $Z$ component's higher sensitivity to Earth's interior motion~\cite{zeng1998manual, ding2004geomagnetic}.
To examine this correlation, we smooth the time sequence in Fig.~\ref{fig:contour}(d) using a Savitzky--Golay filter~\cite{savitzky1964smoothing} with a 30-day window, shown in Fig.~\ref{fig:earthquake}.
We then mark earthquakes (with epicenters in China) of magnitude greater than 6.0, indicated by dashed lines—red for Case~1 ($M \ge 6.5$) and green for Case~2 ($6.0 \le M < 6.5$).

These earthquakes generally occur when $\delta B_z$ approaches its peak, suggesting a potential link between seismic activity and geomagnetic variations. 
Events that do not follow this empirical pattern are marked with gray crosses, and those occurring midway between peaks and valleys are marked with gray hollow circles. 
Overall, about $90\%$ of Case~1 events and $73\%$ of Case~2 events coincide with peak values, supporting this possible correlation.

Earthquakes typically occur over a few seconds, while observatories in the dataset record data at minute-scale intervals. As a result, the short-term variations observed are predominantly shaped by solar effects.
The correlation shown in Fig.~\ref{fig:earthquake} therefore directly links large earthquakes with solar-induced geomagnetic disturbances.
This conclusion is supported by recent satellite observations suggesting a correlation 
between solar activity and large earthquakes~\cite{marchitelli2020correlation}. 
These findings imply that solar activity may contribute energy to triggering earthquakes~\cite{simpson1967solar}.

We interpret these correlations in terms of an energy cascade~\cite{kolmogorov1995turbulence}, which describes the transfer 
of energy from large scales—where it is injected—down to small scales—where it is dissipated. 
In our case, external energy injection occurs via solar events such as the solar wind, which operates on very large scales. This energy is then cascaded to smaller scales, including its surrounding geomagnetic field, and further down to scales smaller than Earth where it is ultimately dissipated through earthquakes. Larger earthquakes require more energy; hence, they tend to occur when $\delta B_z$ is large, indicating a substantial energy transfer. This framework, combined with the observed turbulence-like structure, explains the correlation between earthquakes and solar-induced geomagnetic variations.



{\it Outlooks}
Our study is limited to China, and it remains an open question whether the observed turbulence-like spatial structure of geomagnetic fluctuations is a universal phenomenon across different regions. Future studies should extend this analysis to other geographic locations to assess the generality of our findings.

While our observation of the correlation between the $\delta B_z$ and earthquake is intriguing, it does not constitute a predictive model, as earthquake forecasting also requires knowledge of the spatial location of the epicenter.
A key direction for future work is to explore in greater detail the correlations between the spatial structure of $\delta B_z$ and seismic activity. 
Understanding these relationships could provide new insights into earthquake prediction. 
Additionally, recent advancements in the spatiotemporal prediction of extreme events using machine learning~\cite{qi2020using,jiang2022predicting,pammi2023extreme} and satellite-based earthquake prediction~\cite{zhao2021advances,xiong2021towards,akhoondzadeh2024earthquake} may offer valuable methodologies for this.


\begin{acknowledgments}
The authors thank Suqin Zhang for providing the geomagnetic data. 
\end{acknowledgments}


\begin{thebibliography}{46}%
\makeatletter
\providecommand \@ifxundefined [1]{%
 \@ifx{#1\undefined}
}%
\providecommand \@ifnum [1]{%
 \ifnum #1\expandafter \@firstoftwo
 \else \expandafter \@secondoftwo
 \fi
}%
\providecommand \@ifx [1]{%
 \ifx #1\expandafter \@firstoftwo
 \else \expandafter \@secondoftwo
 \fi
}%
\providecommand \natexlab [1]{#1}%
\providecommand \enquote  [1]{``#1''}%
\providecommand \bibnamefont  [1]{#1}%
\providecommand \bibfnamefont [1]{#1}%
\providecommand \citenamefont [1]{#1}%
\providecommand \href@noop [0]{\@secondoftwo}%
\providecommand \href [0]{\begingroup \@sanitize@url \@href}%
\providecommand \@href[1]{\@@startlink{#1}\@@href}%
\providecommand \@@href[1]{\endgroup#1\@@endlink}%
\providecommand \@sanitize@url [0]{\catcode `\\12\catcode `\$12\catcode `\&12\catcode `\#12\catcode `\^12\catcode `\_12\catcode `\%12\relax}%
\providecommand \@@startlink[1]{}%
\providecommand \@@endlink[0]{}%
\providecommand \url  [0]{\begingroup\@sanitize@url \@url }%
\providecommand \@url [1]{\endgroup\@href {#1}{\urlprefix }}%
\providecommand \urlprefix  [0]{URL }%
\providecommand \Eprint [0]{\href }%
\providecommand \doibase [0]{https://doi.org/}%
\providecommand \selectlanguage [0]{\@gobble}%
\providecommand \bibinfo  [0]{\@secondoftwo}%
\providecommand \bibfield  [0]{\@secondoftwo}%
\providecommand \translation [1]{[#1]}%
\providecommand \BibitemOpen [0]{}%
\providecommand \bibitemStop [0]{}%
\providecommand \bibitemNoStop [0]{.\EOS\space}%
\providecommand \EOS [0]{\spacefactor3000\relax}%
\providecommand \BibitemShut  [1]{\csname bibitem#1\endcsname}%
\let\auto@bib@innerbib\@empty
\bibitem [{\citenamefont {Kivelson}\ and\ \citenamefont {Russell}(1995)}]{kivelson1995introduction}%
  \BibitemOpen
  \bibfield  {author} {\bibinfo {author} {\bibfnamefont {M.~G.}\ \bibnamefont {Kivelson}}\ and\ \bibinfo {author} {\bibfnamefont {C.~T.}\ \bibnamefont {Russell}},\ }\href@noop {} {\emph {\bibinfo {title} {Introduction to space physics}}}\ (\bibinfo  {publisher} {Cambridge university press},\ \bibinfo {year} {1995})\BibitemShut {NoStop}%
\bibitem [{\citenamefont {Basavaiah}(2012)}]{basavaiah2012geomagnetism}%
  \BibitemOpen
  \bibfield  {author} {\bibinfo {author} {\bibfnamefont {N.}~\bibnamefont {Basavaiah}},\ }\href@noop {} {\emph {\bibinfo {title} {Geomagnetism: solid earth and upper atmosphere perspectives}}}\ (\bibinfo  {publisher} {Springer Science \& Business Media},\ \bibinfo {year} {2012})\BibitemShut {NoStop}%
\bibitem [{\citenamefont {Mandea}\ \emph {et~al.}(2019)\citenamefont {Mandea}, \citenamefont {Korte}, \citenamefont {Yau},\ and\ \citenamefont {Petrovsky}}]{mandea2019geomagnetism}%
  \BibitemOpen
  \bibfield  {author} {\bibinfo {author} {\bibfnamefont {M.}~\bibnamefont {Mandea}}, \bibinfo {author} {\bibfnamefont {M.}~\bibnamefont {Korte}}, \bibinfo {author} {\bibfnamefont {A.}~\bibnamefont {Yau}},\ and\ \bibinfo {author} {\bibfnamefont {E.}~\bibnamefont {Petrovsky}},\ }\href@noop {} {\emph {\bibinfo {title} {Geomagnetism, Aeronomy and Space Weather: A Journey from the Earth's Core to the Sun}}},\ Vol.~\bibinfo {volume} {4}\ (\bibinfo  {publisher} {Cambridge University Press},\ \bibinfo {year} {2019})\BibitemShut {NoStop}%
\bibitem [{\citenamefont {Stacey}(1964)}]{stacey1964seismomagnetic}%
  \BibitemOpen
  \bibfield  {author} {\bibinfo {author} {\bibfnamefont {F.~D.}\ \bibnamefont {Stacey}},\ }\href@noop {} {\bibfield  {journal} {\bibinfo  {journal} {Pure and applied Geophysics}\ }\textbf {\bibinfo {volume} {58}},\ \bibinfo {pages} {5} (\bibinfo {year} {1964})}\BibitemShut {NoStop}%
\bibitem [{\citenamefont {Stacey}\ and\ \citenamefont {Johnston}(1972)}]{stacey1972theory}%
  \BibitemOpen
  \bibfield  {author} {\bibinfo {author} {\bibfnamefont {F.~D.}\ \bibnamefont {Stacey}}\ and\ \bibinfo {author} {\bibfnamefont {M.~J.}\ \bibnamefont {Johnston}},\ }\href@noop {} {\bibfield  {journal} {\bibinfo  {journal} {pure and applied geophysics}\ }\textbf {\bibinfo {volume} {97}},\ \bibinfo {pages} {146} (\bibinfo {year} {1972})}\BibitemShut {NoStop}%
\bibitem [{\citenamefont {Johnston}(2002)}]{johnston2002electromagnetic}%
  \BibitemOpen
  \bibfield  {author} {\bibinfo {author} {\bibfnamefont {M.}~\bibnamefont {Johnston}},\ }in\ \href@noop {} {\emph {\bibinfo {booktitle} {International Geophysics}}},\ Vol.~\bibinfo {volume} {81}\ (\bibinfo  {publisher} {Elsevier},\ \bibinfo {year} {2002})\ pp.\ \bibinfo {pages} {621--635}\BibitemShut {NoStop}%
\bibitem [{\citenamefont {Fitterman}(1979)}]{fitterman1979theory}%
  \BibitemOpen
  \bibfield  {author} {\bibinfo {author} {\bibfnamefont {D.~V.}\ \bibnamefont {Fitterman}},\ }\href@noop {} {\bibfield  {journal} {\bibinfo  {journal} {Journal of Geophysical Research: Solid Earth}\ }\textbf {\bibinfo {volume} {84}},\ \bibinfo {pages} {6031} (\bibinfo {year} {1979})}\BibitemShut {NoStop}%
\bibitem [{\citenamefont {Zeng}\ \emph {et~al.}(1998)\citenamefont {Zeng}, \citenamefont {Lin}, \citenamefont {Xu}, \citenamefont {Zhao},\ and\ \citenamefont {Zhao}}]{zeng1998manual}%
  \BibitemOpen
  \bibfield  {author} {\bibinfo {author} {\bibfnamefont {X.}~\bibnamefont {Zeng}}, \bibinfo {author} {\bibfnamefont {Y.}~\bibnamefont {Lin}}, \bibinfo {author} {\bibfnamefont {C.}~\bibnamefont {Xu}}, \bibinfo {author} {\bibfnamefont {M.}~\bibnamefont {Zhao}},\ and\ \bibinfo {author} {\bibfnamefont {Y.}~\bibnamefont {Zhao}},\ }in\ \href@noop {} {\emph {\bibinfo {booktitle} {UN International Workshop on Geomagnetic Methods, Beijing, China}}}\ (\bibinfo {year} {1998})\ pp.\ \bibinfo {pages} {12--18}\BibitemShut {NoStop}%
\bibitem [{\citenamefont {Ding}\ \emph {et~al.}(2004)\citenamefont {Ding}, \citenamefont {Liu}, \citenamefont {Yu},\ and\ \citenamefont {Xiao}}]{ding2004geomagnetic}%
  \BibitemOpen
  \bibfield  {author} {\bibinfo {author} {\bibfnamefont {J.-h.}\ \bibnamefont {Ding}}, \bibinfo {author} {\bibfnamefont {J.}~\bibnamefont {Liu}}, \bibinfo {author} {\bibfnamefont {S.-r.}\ \bibnamefont {Yu}},\ and\ \bibinfo {author} {\bibfnamefont {W.-j.}\ \bibnamefont {Xiao}},\ }\href@noop {} {\bibfield  {journal} {\bibinfo  {journal} {Acta Seismologica Sinica}\ }\textbf {\bibinfo {volume} {17}},\ \bibinfo {pages} {85} (\bibinfo {year} {2004})}\BibitemShut {NoStop}%
\bibitem [{\citenamefont {Rabeh}\ \emph {et~al.}(2010)\citenamefont {Rabeh}, \citenamefont {Miranda},\ and\ \citenamefont {Hvozdara}}]{rabeh2010strong}%
  \BibitemOpen
  \bibfield  {author} {\bibinfo {author} {\bibfnamefont {T.}~\bibnamefont {Rabeh}}, \bibinfo {author} {\bibfnamefont {M.}~\bibnamefont {Miranda}},\ and\ \bibinfo {author} {\bibfnamefont {M.}~\bibnamefont {Hvozdara}},\ }\href@noop {} {\bibfield  {journal} {\bibinfo  {journal} {Natural Hazards}\ }\textbf {\bibinfo {volume} {53}},\ \bibinfo {pages} {561} (\bibinfo {year} {2010})}\BibitemShut {NoStop}%
\bibitem [{\citenamefont {Love}\ and\ \citenamefont {Thomas}(2013)}]{love2013insignificant}%
  \BibitemOpen
  \bibfield  {author} {\bibinfo {author} {\bibfnamefont {J.~J.}\ \bibnamefont {Love}}\ and\ \bibinfo {author} {\bibfnamefont {J.~N.}\ \bibnamefont {Thomas}},\ }\href@noop {} {\bibfield  {journal} {\bibinfo  {journal} {Geophysical Research Letters}\ }\textbf {\bibinfo {volume} {40}},\ \bibinfo {pages} {1165} (\bibinfo {year} {2013})}\BibitemShut {NoStop}%
\bibitem [{\citenamefont {Odintsov}\ \emph {et~al.}(2006)\citenamefont {Odintsov}, \citenamefont {Boyarchuk}, \citenamefont {Georgieva}, \citenamefont {Kirov},\ and\ \citenamefont {Atanasov}}]{odintsov2006long}%
  \BibitemOpen
  \bibfield  {author} {\bibinfo {author} {\bibfnamefont {S.}~\bibnamefont {Odintsov}}, \bibinfo {author} {\bibfnamefont {K.}~\bibnamefont {Boyarchuk}}, \bibinfo {author} {\bibfnamefont {K.}~\bibnamefont {Georgieva}}, \bibinfo {author} {\bibfnamefont {B.}~\bibnamefont {Kirov}},\ and\ \bibinfo {author} {\bibfnamefont {D.}~\bibnamefont {Atanasov}},\ }\href@noop {} {\bibfield  {journal} {\bibinfo  {journal} {Physics and Chemistry of the Earth, Parts a/b/c}\ }\textbf {\bibinfo {volume} {31}},\ \bibinfo {pages} {88} (\bibinfo {year} {2006})}\BibitemShut {NoStop}%
\bibitem [{\citenamefont {Huzaimy}\ and\ \citenamefont {Yumoto}(2011)}]{huzaimy2011possible}%
  \BibitemOpen
  \bibfield  {author} {\bibinfo {author} {\bibfnamefont {J.~M.}\ \bibnamefont {Huzaimy}}\ and\ \bibinfo {author} {\bibfnamefont {K.}~\bibnamefont {Yumoto}},\ }in\ \href@noop {} {\emph {\bibinfo {booktitle} {Proceeding of the 2011 IEEE International Conference on Space Science and Communication (IconSpace)}}}\ (\bibinfo {organization} {IEEE},\ \bibinfo {year} {2011})\ pp.\ \bibinfo {pages} {138--141}\BibitemShut {NoStop}%
\bibitem [{\citenamefont {Han}\ \emph {et~al.}(2004)\citenamefont {Han}, \citenamefont {Guo}, \citenamefont {Wu},\ and\ \citenamefont {Ma}}]{han2004possible}%
  \BibitemOpen
  \bibfield  {author} {\bibinfo {author} {\bibfnamefont {Y.}~\bibnamefont {Han}}, \bibinfo {author} {\bibfnamefont {Z.}~\bibnamefont {Guo}}, \bibinfo {author} {\bibfnamefont {J.}~\bibnamefont {Wu}},\ and\ \bibinfo {author} {\bibfnamefont {L.}~\bibnamefont {Ma}},\ }\href@noop {} {\bibfield  {journal} {\bibinfo  {journal} {Science in China Series G: Physics and Astronomy}\ }\textbf {\bibinfo {volume} {47}},\ \bibinfo {pages} {173} (\bibinfo {year} {2004})}\BibitemShut {NoStop}%
\bibitem [{\citenamefont {Gribbin}(1971)}]{gribbin1971relation}%
  \BibitemOpen
  \bibfield  {author} {\bibinfo {author} {\bibfnamefont {J.}~\bibnamefont {Gribbin}},\ }\href@noop {} {\bibfield  {journal} {\bibinfo  {journal} {Science}\ }\textbf {\bibinfo {volume} {173}},\ \bibinfo {pages} {558} (\bibinfo {year} {1971})}\BibitemShut {NoStop}%
\bibitem [{\citenamefont {Marchitelli}\ \emph {et~al.}(2020)\citenamefont {Marchitelli}, \citenamefont {Harabaglia}, \citenamefont {Troise},\ and\ \citenamefont {De~Natale}}]{marchitelli2020correlation}%
  \BibitemOpen
  \bibfield  {author} {\bibinfo {author} {\bibfnamefont {V.}~\bibnamefont {Marchitelli}}, \bibinfo {author} {\bibfnamefont {P.}~\bibnamefont {Harabaglia}}, \bibinfo {author} {\bibfnamefont {C.}~\bibnamefont {Troise}},\ and\ \bibinfo {author} {\bibfnamefont {G.}~\bibnamefont {De~Natale}},\ }\href@noop {} {\bibfield  {journal} {\bibinfo  {journal} {Scientific reports}\ }\textbf {\bibinfo {volume} {10}},\ \bibinfo {pages} {11495} (\bibinfo {year} {2020})}\BibitemShut {NoStop}%
\bibitem [{\citenamefont {Simpson}(1967)}]{simpson1967solar}%
  \BibitemOpen
  \bibfield  {author} {\bibinfo {author} {\bibfnamefont {J.~F.}\ \bibnamefont {Simpson}},\ }\href@noop {} {\bibfield  {journal} {\bibinfo  {journal} {Earth and Planetary Science Letters}\ }\textbf {\bibinfo {volume} {3}},\ \bibinfo {pages} {417} (\bibinfo {year} {1967})}\BibitemShut {NoStop}%
\bibitem [{\citenamefont {Galtier}\ \emph {et~al.}(2020)\citenamefont {Galtier}, \citenamefont {Laurie},\ and\ \citenamefont {Nazarenko}}]{galtier2020plausible}%
  \BibitemOpen
  \bibfield  {author} {\bibinfo {author} {\bibfnamefont {S.}~\bibnamefont {Galtier}}, \bibinfo {author} {\bibfnamefont {J.}~\bibnamefont {Laurie}},\ and\ \bibinfo {author} {\bibfnamefont {S.~V.}\ \bibnamefont {Nazarenko}},\ }\href@noop {} {\bibfield  {journal} {\bibinfo  {journal} {Universe}\ }\textbf {\bibinfo {volume} {6}},\ \bibinfo {pages} {98} (\bibinfo {year} {2020})}\BibitemShut {NoStop}%
\bibitem [{\citenamefont {Galtier}\ and\ \citenamefont {Nazarenko}(2021)}]{GW2021turbulence}%
  \BibitemOpen
  \bibfield  {author} {\bibinfo {author} {\bibfnamefont {S.}~\bibnamefont {Galtier}}\ and\ \bibinfo {author} {\bibfnamefont {S.~V.}\ \bibnamefont {Nazarenko}},\ }\href {https://doi.org/10.1103/PhysRevLett.127.131101} {\bibfield  {journal} {\bibinfo  {journal} {Phys. Rev. Lett.}\ }\textbf {\bibinfo {volume} {127}},\ \bibinfo {pages} {131101} (\bibinfo {year} {2021})}\BibitemShut {NoStop}%
\bibitem [{\citenamefont {Liang}\ \emph {et~al.}(2025)\citenamefont {Liang}, \citenamefont {Xu}, \citenamefont {Du},\ and\ \citenamefont {Luo}}]{liang2025primordial}%
  \BibitemOpen
  \bibfield  {author} {\bibinfo {author} {\bibfnamefont {J.-x.}\ \bibnamefont {Liang}}, \bibinfo {author} {\bibfnamefont {P.}~\bibnamefont {Xu}}, \bibinfo {author} {\bibfnamefont {M.-h.}\ \bibnamefont {Du}},\ and\ \bibinfo {author} {\bibfnamefont {Z.-r.}\ \bibnamefont {Luo}},\ }\href@noop {} {\bibfield  {journal} {\bibinfo  {journal} {arXiv preprint arXiv:2501.10158}\ } (\bibinfo {year} {2025})}\BibitemShut {NoStop}%
\bibitem [{\citenamefont {Kolmogorov}()}]{kolmogorov1995turbulence}%
  \BibitemOpen
  \bibfield  {author} {\bibinfo {author} {\bibfnamefont {A.}~\bibnamefont {Kolmogorov}},\ }\href@noop {} {\emph {\bibinfo {title} {Turbulence: the legacy of AN Kolmogorov}}}\BibitemShut {NoStop}%
\bibitem [{\citenamefont {De~Michelis}\ and\ \citenamefont {Consolini}(2004)}]{de2004time}%
  \BibitemOpen
  \bibfield  {author} {\bibinfo {author} {\bibfnamefont {P.}~\bibnamefont {De~Michelis}}\ and\ \bibinfo {author} {\bibfnamefont {G.}~\bibnamefont {Consolini}},\ }\href@noop {} {\  (\bibinfo {year} {2004})}\BibitemShut {NoStop}%
\bibitem [{\citenamefont {Borovsky}\ and\ \citenamefont {Funsten}(2003)}]{borovsky2003role}%
  \BibitemOpen
  \bibfield  {author} {\bibinfo {author} {\bibfnamefont {J.~E.}\ \bibnamefont {Borovsky}}\ and\ \bibinfo {author} {\bibfnamefont {H.~O.}\ \bibnamefont {Funsten}},\ }\href@noop {} {\bibfield  {journal} {\bibinfo  {journal} {Journal of Geophysical Research: Space Physics}\ }\textbf {\bibinfo {volume} {108}} (\bibinfo {year} {2003})}\BibitemShut {NoStop}%
\bibitem [{\citenamefont {Zimbardo}(2006)}]{zimbardo2006magnetic}%
  \BibitemOpen
  \bibfield  {author} {\bibinfo {author} {\bibfnamefont {G.}~\bibnamefont {Zimbardo}},\ }\href@noop {} {\bibfield  {journal} {\bibinfo  {journal} {Plasma physics and controlled fusion}\ }\textbf {\bibinfo {volume} {48}},\ \bibinfo {pages} {B295} (\bibinfo {year} {2006})}\BibitemShut {NoStop}%
\bibitem [{\citenamefont {Benzi}\ \emph {et~al.}(1993{\natexlab{a}})\citenamefont {Benzi}, \citenamefont {Ciliberto}, \citenamefont {Tripiccione}, \citenamefont {Baudet}, \citenamefont {Massaioli},\ and\ \citenamefont {Succi}}]{benzi1993extended}%
  \BibitemOpen
  \bibfield  {author} {\bibinfo {author} {\bibfnamefont {R.}~\bibnamefont {Benzi}}, \bibinfo {author} {\bibfnamefont {S.}~\bibnamefont {Ciliberto}}, \bibinfo {author} {\bibfnamefont {R.}~\bibnamefont {Tripiccione}}, \bibinfo {author} {\bibfnamefont {C.}~\bibnamefont {Baudet}}, \bibinfo {author} {\bibfnamefont {F.}~\bibnamefont {Massaioli}},\ and\ \bibinfo {author} {\bibfnamefont {S.}~\bibnamefont {Succi}},\ }\href@noop {} {\bibfield  {journal} {\bibinfo  {journal} {Physical Review E}\ }\textbf {\bibinfo {volume} {48}},\ \bibinfo {pages} {R29} (\bibinfo {year} {1993}{\natexlab{a}})}\BibitemShut {NoStop}%
\bibitem [{\citenamefont {Van~Brummelen}(2017)}]{van2017heavenly}%
  \BibitemOpen
  \bibfield  {author} {\bibinfo {author} {\bibfnamefont {G.}~\bibnamefont {Van~Brummelen}},\ }\href@noop {} {\emph {\bibinfo {title} {Heavenly mathematics: The forgotten art of spherical trigonometry}}}\ (\bibinfo  {publisher} {Princeton University Press},\ \bibinfo {year} {2017})\BibitemShut {NoStop}%
\bibitem [{\citenamefont {Aluie}(2011)}]{aluie2011compressible}%
  \BibitemOpen
  \bibfield  {author} {\bibinfo {author} {\bibfnamefont {H.}~\bibnamefont {Aluie}},\ }\href@noop {} {\bibfield  {journal} {\bibinfo  {journal} {Physical review letters}\ }\textbf {\bibinfo {volume} {106}},\ \bibinfo {pages} {174502} (\bibinfo {year} {2011})}\BibitemShut {NoStop}%
\bibitem [{\citenamefont {Aluie}(2013)}]{aluie2013scale}%
  \BibitemOpen
  \bibfield  {author} {\bibinfo {author} {\bibfnamefont {H.}~\bibnamefont {Aluie}},\ }\href@noop {} {\bibfield  {journal} {\bibinfo  {journal} {Physica D: Nonlinear Phenomena}\ }\textbf {\bibinfo {volume} {247}},\ \bibinfo {pages} {54} (\bibinfo {year} {2013})}\BibitemShut {NoStop}%
\bibitem [{\citenamefont {Kolmogorov}(1991{\natexlab{a}})}]{kolmogorov1991local}%
  \BibitemOpen
  \bibfield  {author} {\bibinfo {author} {\bibfnamefont {A.~N.}\ \bibnamefont {Kolmogorov}},\ }\href@noop {} {\bibfield  {journal} {\bibinfo  {journal} {Proceedings of the Royal Society of London. Series A: Mathematical and Physical Sciences}\ }\textbf {\bibinfo {volume} {434}},\ \bibinfo {pages} {9} (\bibinfo {year} {1991}{\natexlab{a}})}\BibitemShut {NoStop}%
\bibitem [{\citenamefont {Kolmogorov}(1941)}]{kolmogorov1941degeneration}%
  \BibitemOpen
  \bibfield  {author} {\bibinfo {author} {\bibfnamefont {A.~N.}\ \bibnamefont {Kolmogorov}},\ }in\ \href@noop {} {\emph {\bibinfo {booktitle} {Dokl. Akad. Nauk SSSR}}},\ Vol.~\bibinfo {volume} {31}\ (\bibinfo {year} {1941})\ pp.\ \bibinfo {pages} {538--540}\BibitemShut {NoStop}%
\bibitem [{\citenamefont {Kolmogorov}(1991{\natexlab{b}})}]{kolmogorov1991dissipation}%
  \BibitemOpen
  \bibfield  {author} {\bibinfo {author} {\bibfnamefont {A.~N.}\ \bibnamefont {Kolmogorov}},\ }\href@noop {} {\bibfield  {journal} {\bibinfo  {journal} {Proceedings of the Royal Society of London. Series A: Mathematical and Physical Sciences}\ }\textbf {\bibinfo {volume} {434}},\ \bibinfo {pages} {15} (\bibinfo {year} {1991}{\natexlab{b}})}\BibitemShut {NoStop}%
\bibitem [{\citenamefont {Benzi}\ \emph {et~al.}(1993{\natexlab{b}})\citenamefont {Benzi}, \citenamefont {Ciliberto}, \citenamefont {Baudet}, \citenamefont {Chavarria},\ and\ \citenamefont {Tripiccione}}]{benzi1993extended2}%
  \BibitemOpen
  \bibfield  {author} {\bibinfo {author} {\bibfnamefont {R.}~\bibnamefont {Benzi}}, \bibinfo {author} {\bibfnamefont {S.}~\bibnamefont {Ciliberto}}, \bibinfo {author} {\bibfnamefont {C.}~\bibnamefont {Baudet}}, \bibinfo {author} {\bibfnamefont {G.~R.}\ \bibnamefont {Chavarria}},\ and\ \bibinfo {author} {\bibfnamefont {R.}~\bibnamefont {Tripiccione}},\ }\href@noop {} {\bibfield  {journal} {\bibinfo  {journal} {Europhysics letters}\ }\textbf {\bibinfo {volume} {24}},\ \bibinfo {pages} {275} (\bibinfo {year} {1993}{\natexlab{b}})}\BibitemShut {NoStop}%
\bibitem [{\citenamefont {Carbone}\ \emph {et~al.}(1996)\citenamefont {Carbone}, \citenamefont {Bruno},\ and\ \citenamefont {Veltri}}]{carbone1996evidences}%
  \BibitemOpen
  \bibfield  {author} {\bibinfo {author} {\bibfnamefont {V.}~\bibnamefont {Carbone}}, \bibinfo {author} {\bibfnamefont {R.}~\bibnamefont {Bruno}},\ and\ \bibinfo {author} {\bibfnamefont {P.}~\bibnamefont {Veltri}},\ }\href@noop {} {\bibfield  {journal} {\bibinfo  {journal} {Geophysical research letters}\ }\textbf {\bibinfo {volume} {23}},\ \bibinfo {pages} {121} (\bibinfo {year} {1996})}\BibitemShut {NoStop}%
\bibitem [{\citenamefont {Nikora}\ and\ \citenamefont {Goring}(2001)}]{nikora2001extended}%
  \BibitemOpen
  \bibfield  {author} {\bibinfo {author} {\bibfnamefont {V.~I.}\ \bibnamefont {Nikora}}\ and\ \bibinfo {author} {\bibfnamefont {D.~G.}\ \bibnamefont {Goring}},\ }\href@noop {} {\bibfield  {journal} {\bibinfo  {journal} {Mathematical geology}\ }\textbf {\bibinfo {volume} {33}},\ \bibinfo {pages} {251} (\bibinfo {year} {2001})}\BibitemShut {NoStop}%
\bibitem [{\citenamefont {Kolmogorov}(1962)}]{kolmogorov1962refinement}%
  \BibitemOpen
  \bibfield  {author} {\bibinfo {author} {\bibfnamefont {A.~N.}\ \bibnamefont {Kolmogorov}},\ }\href@noop {} {\bibfield  {journal} {\bibinfo  {journal} {Journal of Fluid Mechanics}\ }\textbf {\bibinfo {volume} {13}},\ \bibinfo {pages} {82} (\bibinfo {year} {1962})}\BibitemShut {NoStop}%
\bibitem [{\citenamefont {Obukhov}(1962)}]{obukhov1962some}%
  \BibitemOpen
  \bibfield  {author} {\bibinfo {author} {\bibfnamefont {A.}~\bibnamefont {Obukhov}},\ }\href@noop {} {\bibfield  {journal} {\bibinfo  {journal} {Journal of Geophysical Research}\ }\textbf {\bibinfo {volume} {67}},\ \bibinfo {pages} {3011} (\bibinfo {year} {1962})}\BibitemShut {NoStop}%
\bibitem [{\citenamefont {Ching}\ \emph {et~al.}(2002)\citenamefont {Ching}, \citenamefont {She}, \citenamefont {Su},\ and\ \citenamefont {Zou}}]{ching2002extended}%
  \BibitemOpen
  \bibfield  {author} {\bibinfo {author} {\bibfnamefont {E.~S.}\ \bibnamefont {Ching}}, \bibinfo {author} {\bibfnamefont {Z.-S.}\ \bibnamefont {She}}, \bibinfo {author} {\bibfnamefont {W.}~\bibnamefont {Su}},\ and\ \bibinfo {author} {\bibfnamefont {Z.}~\bibnamefont {Zou}},\ }\href@noop {} {\bibfield  {journal} {\bibinfo  {journal} {Physical Review E}\ }\textbf {\bibinfo {volume} {65}},\ \bibinfo {pages} {066303} (\bibinfo {year} {2002})}\BibitemShut {NoStop}%
\bibitem [{\citenamefont {Toschi}\ \emph {et~al.}(1999)\citenamefont {Toschi}, \citenamefont {Amati}, \citenamefont {Succi}, \citenamefont {Benzi},\ and\ \citenamefont {Piva}}]{toschi1999intermittency}%
  \BibitemOpen
  \bibfield  {author} {\bibinfo {author} {\bibfnamefont {F.}~\bibnamefont {Toschi}}, \bibinfo {author} {\bibfnamefont {G.}~\bibnamefont {Amati}}, \bibinfo {author} {\bibfnamefont {S.}~\bibnamefont {Succi}}, \bibinfo {author} {\bibfnamefont {R.}~\bibnamefont {Benzi}},\ and\ \bibinfo {author} {\bibfnamefont {R.}~\bibnamefont {Piva}},\ }\href@noop {} {\bibfield  {journal} {\bibinfo  {journal} {Physical review letters}\ }\textbf {\bibinfo {volume} {82}},\ \bibinfo {pages} {5044} (\bibinfo {year} {1999})}\BibitemShut {NoStop}%
\bibitem [{\citenamefont {Arad}\ \emph {et~al.}(1999)\citenamefont {Arad}, \citenamefont {Biferale}, \citenamefont {Mazzitelli},\ and\ \citenamefont {Procaccia}}]{arad1999disentangling}%
  \BibitemOpen
  \bibfield  {author} {\bibinfo {author} {\bibfnamefont {I.}~\bibnamefont {Arad}}, \bibinfo {author} {\bibfnamefont {L.}~\bibnamefont {Biferale}}, \bibinfo {author} {\bibfnamefont {I.}~\bibnamefont {Mazzitelli}},\ and\ \bibinfo {author} {\bibfnamefont {I.}~\bibnamefont {Procaccia}},\ }\href@noop {} {\bibfield  {journal} {\bibinfo  {journal} {Physical review letters}\ }\textbf {\bibinfo {volume} {82}},\ \bibinfo {pages} {5040} (\bibinfo {year} {1999})}\BibitemShut {NoStop}%
\bibitem [{\citenamefont {Savitzky}\ and\ \citenamefont {Golay}(1964)}]{savitzky1964smoothing}%
  \BibitemOpen
  \bibfield  {author} {\bibinfo {author} {\bibfnamefont {A.}~\bibnamefont {Savitzky}}\ and\ \bibinfo {author} {\bibfnamefont {M.~J.}\ \bibnamefont {Golay}},\ }\href@noop {} {\bibfield  {journal} {\bibinfo  {journal} {Analytical chemistry}\ }\textbf {\bibinfo {volume} {36}},\ \bibinfo {pages} {1627} (\bibinfo {year} {1964})}\BibitemShut {NoStop}%
\bibitem [{\citenamefont {Qi}\ and\ \citenamefont {Majda}(2020)}]{qi2020using}%
  \BibitemOpen
  \bibfield  {author} {\bibinfo {author} {\bibfnamefont {D.}~\bibnamefont {Qi}}\ and\ \bibinfo {author} {\bibfnamefont {A.~J.}\ \bibnamefont {Majda}},\ }\href@noop {} {\bibfield  {journal} {\bibinfo  {journal} {Proceedings of the National Academy of Sciences}\ }\textbf {\bibinfo {volume} {117}},\ \bibinfo {pages} {52} (\bibinfo {year} {2020})}\BibitemShut {NoStop}%
\bibitem [{\citenamefont {Jiang}\ \emph {et~al.}(2022)\citenamefont {Jiang}, \citenamefont {Huang}, \citenamefont {Grebogi},\ and\ \citenamefont {Lai}}]{jiang2022predicting}%
  \BibitemOpen
  \bibfield  {author} {\bibinfo {author} {\bibfnamefont {J.}~\bibnamefont {Jiang}}, \bibinfo {author} {\bibfnamefont {Z.-G.}\ \bibnamefont {Huang}}, \bibinfo {author} {\bibfnamefont {C.}~\bibnamefont {Grebogi}},\ and\ \bibinfo {author} {\bibfnamefont {Y.-C.}\ \bibnamefont {Lai}},\ }\href@noop {} {\bibfield  {journal} {\bibinfo  {journal} {Physical Review Research}\ }\textbf {\bibinfo {volume} {4}},\ \bibinfo {pages} {023028} (\bibinfo {year} {2022})}\BibitemShut {NoStop}%
\bibitem [{\citenamefont {Pammi}\ \emph {et~al.}(2023)\citenamefont {Pammi}, \citenamefont {Clerc}, \citenamefont {Coulibaly},\ and\ \citenamefont {Barbay}}]{pammi2023extreme}%
  \BibitemOpen
  \bibfield  {author} {\bibinfo {author} {\bibfnamefont {V.}~\bibnamefont {Pammi}}, \bibinfo {author} {\bibfnamefont {M.}~\bibnamefont {Clerc}}, \bibinfo {author} {\bibfnamefont {S.}~\bibnamefont {Coulibaly}},\ and\ \bibinfo {author} {\bibfnamefont {S.}~\bibnamefont {Barbay}},\ }\href@noop {} {\bibfield  {journal} {\bibinfo  {journal} {Physical Review Letters}\ }\textbf {\bibinfo {volume} {130}},\ \bibinfo {pages} {223801} (\bibinfo {year} {2023})}\BibitemShut {NoStop}%
\bibitem [{\citenamefont {Zhao}\ \emph {et~al.}(2021)\citenamefont {Zhao}, \citenamefont {Pan}, \citenamefont {Sun}, \citenamefont {Guo}, \citenamefont {Zhang},\ and\ \citenamefont {Feng}}]{zhao2021advances}%
  \BibitemOpen
  \bibfield  {author} {\bibinfo {author} {\bibfnamefont {X.}~\bibnamefont {Zhao}}, \bibinfo {author} {\bibfnamefont {S.}~\bibnamefont {Pan}}, \bibinfo {author} {\bibfnamefont {Z.}~\bibnamefont {Sun}}, \bibinfo {author} {\bibfnamefont {H.}~\bibnamefont {Guo}}, \bibinfo {author} {\bibfnamefont {L.}~\bibnamefont {Zhang}},\ and\ \bibinfo {author} {\bibfnamefont {K.}~\bibnamefont {Feng}},\ }\href@noop {} {\bibfield  {journal} {\bibinfo  {journal} {Natural Hazards Review}\ }\textbf {\bibinfo {volume} {22}},\ \bibinfo {pages} {03120001} (\bibinfo {year} {2021})}\BibitemShut {NoStop}%
\bibitem [{\citenamefont {Xiong}\ \emph {et~al.}(2021)\citenamefont {Xiong}, \citenamefont {Tong}, \citenamefont {Zhang}, \citenamefont {Shen}, \citenamefont {Battiston}, \citenamefont {Ouzounov}, \citenamefont {Iuppa}, \citenamefont {Crookes}, \citenamefont {Long},\ and\ \citenamefont {Zhou}}]{xiong2021towards}%
  \BibitemOpen
  \bibfield  {author} {\bibinfo {author} {\bibfnamefont {P.}~\bibnamefont {Xiong}}, \bibinfo {author} {\bibfnamefont {L.}~\bibnamefont {Tong}}, \bibinfo {author} {\bibfnamefont {K.}~\bibnamefont {Zhang}}, \bibinfo {author} {\bibfnamefont {X.}~\bibnamefont {Shen}}, \bibinfo {author} {\bibfnamefont {R.}~\bibnamefont {Battiston}}, \bibinfo {author} {\bibfnamefont {D.}~\bibnamefont {Ouzounov}}, \bibinfo {author} {\bibfnamefont {R.}~\bibnamefont {Iuppa}}, \bibinfo {author} {\bibfnamefont {D.}~\bibnamefont {Crookes}}, \bibinfo {author} {\bibfnamefont {C.}~\bibnamefont {Long}},\ and\ \bibinfo {author} {\bibfnamefont {H.}~\bibnamefont {Zhou}},\ }\href@noop {} {\bibfield  {journal} {\bibinfo  {journal} {Science of the Total Environment}\ }\textbf {\bibinfo {volume} {771}},\ \bibinfo {pages} {145256} (\bibinfo {year} {2021})}\BibitemShut {NoStop}%
\bibitem [{\citenamefont {Akhoondzadeh}(2024)}]{akhoondzadeh2024earthquake}%
  \BibitemOpen
  \bibfield  {author} {\bibinfo {author} {\bibfnamefont {M.}~\bibnamefont {Akhoondzadeh}},\ }\href@noop {} {\bibfield  {journal} {\bibinfo  {journal} {Advances in Space Research}\ }\textbf {\bibinfo {volume} {74}},\ \bibinfo {pages} {3539} (\bibinfo {year} {2024})}\BibitemShut {NoStop}%
\end{thebibliography}
%

\end{document}